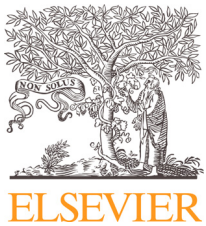

Comment

# Using sensitive data to prevent discrimination by artificial intelligence: Does the GDPR need a new exception?☆

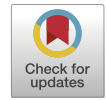

## Marvin van Bekkum\*, Frederik Zuiderveen Borgesius

*Interdisciplinary hub for Digitalization & Society (iHub) & Institute for Computing and Information Sciences (iCIS), Radboud University, Erasmuslaan 1, 6525 GE, Nijmegen, NL, the Netherlands*

ABSTRACT

Organisations can use artificial intelligence to make decisions about people for a variety of reasons, for instance, to select the best candidates from many job applications. However, AI systems can have discriminatory effects when used for decision-making. To illustrate, an AI system could reject applications of people with a certain ethnicity, while the organisation did not plan such ethnicity discrimination. But in Europe, an organisation runs into a problem when it wants to assess whether its AI system accidentally discriminates based on ethnicity: the organisation may not know the applicants' ethnicity. In principle, the GDPR bans the use of certain 'special categories of data' (sometimes called 'sensitive data'), which include data on ethnicity, religion, and sexual preference. The proposal for an AI Act of the European Commission includes a provision that would enable organisations to use special categories of data for auditing their AI systems. This paper asks whether the GDPR's rules on special categories of personal data hinder the prevention of AI-driven discrimination. We argue that the GDPR does prohibit such use of special category data in many circumstances. We also map out the arguments for and against creating an exception to the GDPR's ban on using special categories of personal data, to enable preventing discrimination by AI systems. The paper discusses European law, but the paper can be relevant outside Europe too, as many policymakers in the world grapple with the tension between privacy and non-discrimination policy.

---

☆ Both authors contributed equally to the paper. The authors thank Tom Heskes, Michael Veale, Raphaële Xenidis and Steven Bellovin for their insightful comments and discussion. The authors thank all the participants of the Privacy Law Scholars Conference and Digital Legal Talks for their active discussion.
\* Corresponding author.
*E-mail addresses:* marvin.vanbekkum@ru.nl (M. van Bekkum), frederikzb@cs.ru.nl (F. Zuiderveen Borgesius).
https://doi.org/10.1016/j.clsr.2022.105770

## 1. Introduction

Artificial intelligence (AI) can have discriminatory effects. Suppose that an organisation uses an AI system to decide which candidates to select from many job applications. The AI system rejects job applications from people with certain characteristics, say the educational courses taken by the job applicant. The organisation wants to prevent discrimination based on ethnicity or similar protected grounds.

To assess whether its AI system harms people with a certain ethnicity, the organisation needs to know the ethnicity of its job applicants. In Europe, however, the organisation may not know the ethnicity because, in principle, the GDPR prohibits the use of 'special categories of data' (article 9).[1] Special categories of data include data on ethnicity, religion, health, and sexual preferences. Hence it appears that the organisation is not allowed to infer, collect, or use the ethnicity of the applicants.

Our paper focuses on the following questions. (i) Do the GDPR's rules on special categories of personal data hinder the prevention of AI-driven discrimination? (ii) What are the arguments for and against creating an exception to the GDPR's ban on using special categories of personal data, to enable preventing discrimination by AI systems?

We use the word 'discrimination' mostly to refer to objectionable or illegal discrimination. Hence, we do not use 'discrimination' in a neutral sense. With 'preventing' AI-driven discrimination, we mean detecting and mitigating discrimination by AI systems.[2] An exception to the GDPR's ban on using special categories of data could be included in the GDPR, or in another statute, national or EU-wide.

We focus only on AI systems that have discriminatory effects relating to certain protected grounds in EU non-discrimination directives, namely ethnicity, religion or belief, disability, and sexual orientation.[3] Hence, many types of unfair or discriminatory AI are outside the scope of this paper.

Our paper makes three contributions to the literature. First, we combine legal insights from non-discrimination scholarship on the one hand, and privacy and data protection scholarship on the other hand. We also consider insights from AI and other computer science scholarship. Most existing literature on using special category data for non-discrimination purposes is fragmented between disciplines and sub-disciplines.

Second, we show that, in most circumstances, the GDPR does not allow an organisation to use special categories of data for debiasing. The GDPR allows the EU or national lawmakers to adopt an exception under certain conditions, but lawmakers have not adopted such an exception. We thus disagree with some non-discrimination scholars who appear to suggest that data protection law allows such data collection.[4]

Third, we map out and analyse the arguments for and against introducing a new exception to the GDPR to enable debiasing AI systems. We show that there are valid arguments for both viewpoints.

The paper can be relevant for computer scientists, legal scholars, and policymakers. This paper focuses on the law in Europe. However, the paper can be relevant outside Europe too. The problem of discriminatory AI is high on the agenda of policymakers worldwide.[5] In many countries, privacy law and non-discrimination policy can be in conflict. We aim to describe European law in such a way that non-specialists can follow the discussion too.

The paper is structured as follows. In Sections 2 and 3 we introduce some key terms, summarise how AI can discriminate based on ethnicity and similar characteristics, and we introduce non-discrimination law. In Sections 4 and 5 we analyse the GDPR's framework for special categories of data and we discuss whether the framework hinders organisations in collecting special categories of data. In Section 6 we map out the arguments in favour of and against introducing a new exception to the collection and use of special category data for auditing AI systems. Section 7 discusses possible safeguards that could accompany a new exception, and Section 8 concludes.

---

[1] Also called "special category data" for short.

[2] For more information about the causes of AI-driven discrimination, see Section 2 of this paper.

[3] Racial or ethnic origin: Council Directive 2000/43/EC Implementing the Principle of Equal Treatment Between Persons Irrespective of Racial or Ethnic Origin, 2000 OJ L 180/22. Religion or belief, disability, age, or sexual orientation in the Employment context: Council Directive 2000/78/EC Establishing a General Framework for Equal Treatment in Employment and Occupation, 2000 OJ L 303/16. Gender in the context of the supply of goods and services: Council Directive 2004/113/EC Implementing the principle of Equal Treatment between Men and Women in the Access to and Supply of Goods and Services, 2004 OJ L 373/37. Gender in the employment context: Directive 2006/54/EC of the European Parliament and of the Council on the implementation of the Principle of Equal Opportunities and Equal Treatment of Men and Women in Matters of Employment and Occupation (Recast), 2006 OJ L 204/23. Age and gender are not special categories of data, although they are protected discrimination grounds. See Article 9(1) GDPR.

[4] See Section 5.2.

[5] See e.g. European Commission, COM(2020) 65 *White Paper On Artificial Intelligence - A European Approach to Excellence and Trust* (EU 2020) 1, 9–12, 18–22 <https://ec.europa.eu/info/sites/default/files/commission-white-paper-artificial-intelligence-feb2020_en.pdf>. See also the EU Digital Services Act with requirements for certain types of online platforms to assess discrimination risks. Articles 34(1)(b) and 40(4) of the 'Regulation (EU) 2022/2065 of the European Parliament and of the Council of 19 October 2022 on a Single Market For Digital Services and Amending Directive 2000/31/EC (Digital Services Act)' <https://eur-lex.europa.eu/legal-content/EN/TXT/?uri=CELEX:32022R2065>. In the US: e.g. the White house is planning an 'AI Bill of Rights'. White house employees announced a national effort to develop a "Bill of Rights for an AI-Powered World" in WIRED. See White House Office of Science and Technology Policy (OSTP), 'ICYMI: WIRED (Opinion): Americans Need a Bill of Rights for an AI-Powered World' (22 October 2021) <https://www.whitehouse.gov/ostp/news-updates/2021/10/22/icymi-wired-opinion-americans-need-a-bill-of-rights-for-an-ai-powered-world/> accessed 26 April 2022. A "blueprint" for the AI bill of rights was released: The White House, 'Blueprint for an AI Bill of Rights. Making Automated Systems Work for the American People' <https://www.whitehouse.gov/ostp/ai-bill-of-rights/>. See also 'OECD AI's Live Repository of over 260 AI Strategies & Policies' <https://oecd.ai/en/dashboards> accessed 12 October 2022.

## 2. AI systems and discrimination

AI systems could be described, in the words of the Oxford Dictionary, as 'computer systems able to perform tasks normally requiring human intelligence, such as visual perception, speech recognition, decision-making, and translation between languages.'[6] More technical and complicated definitions exist, which we will not discuss in this paper.[7] We focus on AI systems that make decisions that can have serious effects for people. For example, a bank could use an AI system to decide whether a customer gets a mortgage or not. AI systems can help our society in many ways, and can also help make society fairer. In this paper, however, we focus on one specific risk of AI: the risk that AI systems discriminate against people with certain protected characteristics, such as ethnicity.

### 2.1. Causes of discrimination by AI

Discriminatory input data is one of the main sources of discrimination by AI systems.[8] To illustrate, suppose that an organisation's human resources (HR) personnel has discriminated against women in the past. Let's assume that the organisation does not realise that its HR personnel discriminated in the past. If the organisation uses the historical decisions by humans to train its AI system, the AI system could reproduce that discrimination. Reportedly, a recruitment system developed by Amazon ran into such a problem. Amazon abandoned the project before using it in real recruitment decisions.[9]

AI systems can make discriminatory decisions about job applicants, harming certain ethnicities for instance, even if the system does not have direct access to data about people's ethnicity. Imagine an AI system that considers the postal codes where job applicants live. The postal codes could correlate with someone's ethnicity. Hence, the system might reject all people with a certain ethnicity, even if the organisation has ensured that the system does not consider people's ethnicity. In practice, an AI system might also consider hundreds of variables, in complicated combinations, that turn out to correlate with ethnicity. Variables that correlate with protected attributes such as ethnicity can be called 'proxy attributes'. Such correlations can lead to discrimination by proxy.[10] Because of proxy attributes, AI systems can have discriminatory effects by accident: AI developers or organisations using AI systems may not realise that the AI system discriminates. More causes of discrimination by AI systems exist that we will not delve into further in this paper, ranging from design decisions to the context in which the system is used.[11]

### 2.2. Using special categories of data is useful to debias AI systems

Suppose that an organisation wants to test whether its AI system unfairly discriminates against job applicants with a certain ethnicity. To test this, the organisation must know the ethnicity of both the people who applied for the job, and of the people the organisation actually hired. Say that half of the people who sent in a job application letter has an immigrant background. The AI system selects the fifty best letters, out of the thousands of letters. The AI system decides based on attributes such as the school of choice, or courses followed. Of the fifty letters selected by the AI system, none is by somebody with an immigrant background. Such numbers suggest that the AI system should be investigated for unfair or illegal bias. Because proxy attributes may hide discrimination, the special categories of data are necessary for a detailed analysis.[12] In sum, collecting special categories of data (such as ethnicity data) is often useful, or even necessary, to audit AI systems for discrimination.[13]

## 3. Non-discrimination law

The right to non-discrimination is a human right. It is protected, for instance, in the European Convention on Human Rights (1950),[14] the International Convention on the Elimination of all Forms of Racial Discrimination (1965),[15] the International Covenant on Civil and Political Rights (1966),[16] and

---

[6] 'Artificial Intelligence' (*Oxford Reference*) <https://www.oxfordreference.com/view/10.1093/oi/authority.20110803095426960> accessed 12 October 2022.

[7] See generally on defining AI: High-level expert group on Artificial Intelligence, *A Definition of AI. Main Capabilities and Disciplines* (European Commission 2019) <https://ec.europa.eu/newsroom/dae/redirection/document/56341>.

[8] There are other possible causes. See for an overview of ways in which AI systems can have discriminatory effects: S Barocas and A Selbst, 'Big Data's Disparate Impact' [2016] 104 Calif Law Rev 671 <https://www.jstor.org/stable/24758720>.;F Zuiderveen Borgesius, *Discrimination, Artificial Intelligence, and Algorithmic Decision-Making. Report for the European Commission against Racism and Intolerance (ECRI)* (Council of Europe 2019) s III.2 <https://www.coe.int/en/web/artificial-intelligence/-/news-of-the-european-commission-against-racism-and-intolerance-ecri->.
A possible classification of biases was created by TILT. Tilburg Institute for Law, Technology, and Society, *Handbook on Non-Discriminating Algorithms. Summary Research Report* (2021) 5 <https://www.tilburguniversity.edu/about/schools/law/departments/tilt/research/handbook>.

[9] A recruitment system used by Amazon until October 2018 seemed to display this type of bias. See Reuters, 'Amazon Scraps Secret AI Recruiting Tool That Showed Bias against Women' (11 October 2018) <https://www.reuters.com/article/us-amazon-com-jobs-automation-insight-idUSKCN1MK08G> accessed 7 April 2022.

[10] Barocas and Selbst (n 8) 675 & 712. I Žliobaitė and B Custers, 'Using Sensitive Personal Data May Be Necessary for Avoiding Discrimination in Data-Driven Decision Models' (2016) 24 Artificial Intelligence and Law 183, 185 <http://link.springer.com/10.1007/s10506-016-9182-5>.

[11] See in-depth Barocas and Selbst (n 8).

[12] See more in-depth Žliobaitė and Custers (n 10) 190–193.

[13] There are practical hurdles to testing an AI-system for discrimination. See the end of Section 6.3 of this paper.

[14] Article 14 of the European Convention on Human Rights.

[15] See in particular article 1-7 of the International Convention on the Elimination of all Forms of Racial Discrimination.

[16] Article 2 and 26 of the International Covenant on Civil and Political Rights.

the Charter of Fundamental Rights of the European Union (2000).[17]

Here, we focus on European Union law. EU law forbids two forms of discrimination: direct and indirect discrimination. The Racial Equality Directive (about ethnicity) states that 'the principle of equal treatment shall mean that there shall be no direct or indirect discrimination based on racial or ethnic origin.'[18] The Racial Equality Directive describes *direct* discrimination as follows: 'Direct discrimination shall be taken to occur where one person is treated less favourably than another is, has been or would be treated in a comparable situation on grounds of racial or ethnic origin.'[19] A clear example of direct discrimination is the discrimination against Black South Africans by the Apartheid regime in South-Africa in the 20th century.[20]

Direct discrimination is prohibited in EU law. There are some specific, narrowly defined, exceptions to this prohibition. For example, the Racial Equality Directive allows a difference in treatment based on ethnicity if ethnicity 'constitutes a genuine and determining occupational requirement'.[21] A fitting example is the choice of a black actor to play the part of Othello.[22]

In non-discrimination scholarship, the grounds such as ethnicity ('racial or ethnic origin') are called 'protected characteristics' or 'protected attributes'. An AI system that treats individuals differently based on their protected attributes would discriminate directly. A hypothetical example of direct discrimination by a computer system is if the programmer explicitly makes the system reject all women.

Our paper, however, focuses on indirect discrimination. The Racial Equality Directive defines indirect discrimination as follows.

> [I]ndirect discrimination shall be taken to occur where (i) an apparently neutral provision, criterion or practice would (ii) put persons of a racial or ethnic origin at a particular disadvantage compared with other persons, (iii) unless that provision, criterion or practice is objectively justified by a legitimate aim and the means of achieving that aim are appropriate and necessary.[23]

In short, indirect discrimination by an AI system can occur if the system ('practice') is neutral at first glance but ends up discriminating against people with a protected characteristic. For example, even if the protected attributes in an AI system were filtered out, the system can still discriminate based on metrics that are a proxy for the protected attribute. Say that a recruitment system bases its decision on the set of courses, age, and origin university from the job applicant's CV. Such metrics could correlate with ethnicity or another protected attribute.

For both direct and indirect discrimination, it is irrelevant whether the organisation discriminates by accident or on purpose. Hence, an organisation is always liable, even if the organisation did not realise that its AI system was indirectly discriminating.[24]

Unlike for direct discrimination, for indirect discrimination there is an open-ended exception. Indirect discrimination is allowed if there is an objective justification. The possibility for justification is part of the definition of indirect discrimination: 'unless that provision, criterion or practice is objectively justified'.[25] In short, if the organisation (the alleged discriminator) has a legitimate aim for its neutral practice, and that practice is a proportional way of aiming for that practice, there is no illegal indirect discrimination.[26] If an AI system has discriminatory effects, the general norms from EU non-discrimination law apply.

## 4. Data protection law

The right to privacy and the right to the protection of personal data are both fundamental rights. Privacy is protected, for instance, in the European Convention on Human Rights (1950),[27] the International Covenant on Civil and Political Rights (1966),[28] and the Charter of Fundamental Rights of the European Union (2000).[29]

Since the 1970s, a new field of law has been developed: data protection law. In the EU, the right to the protection of personal data has the status of a fundamental right.[30] The right to pro-tection of personal data is explicitly protected in the Charter of Fundamental Rights of the European Union.[31] Data protection law grants rights to people whose data are being processed (data subjects), and imposes obligations on parties that process personal data (data controllers). Data protection law aims to protect personal data, and in doing so protects other values and rights. Unlike some seem to assume, data protection law does not only aim to protect privacy, but also aims to protect the right to non-discrimination and other rights.

---

[17] Article 21 of the Charter of Fundamental Rights of the European Union. See also article 19 of the Treaty on the Functioning of the European Union.
[18] See Article 2 Council Directive 2000/43/EC Implementing the Principle of Equal Treatment Between Persons Irrespective of Racial or Ethnic Origin, 2000 OJ L 180/22.
[19] Article 2(2)(a) Racial Equality Directive 2000/43/EC.
[20] See for example FK Thomsen, 'Direct Discrimination' in K Lippert-Rasmussen, *The Routledge Handbook of the Ethics of Discrimination* (Routledge Taylor & Francis Group 2018).
[21] Article 4 Racial Equality Directive 2000/43/EC.
[22] E Ellis and P Watson, *EU Anti-Discrimination Law* (Oxford University Press 2012) 382.
[23] Article 2(2)(b) of the Racial Equality Directive 2000/43/EC. Capitalisation and interpunction amended by the authors.
[24] For completeness' sake, we add that the difference between direct and indirect discrimination can be somewhat fuzzy. That discussion, however, falls outside the scope of this paper. See e.g. European Union Agency for Fundamental Rights, European Court of Human Rights, and Council of Europe (Strasbourg), *Handbook on European Non-Discrimination Law* (Publications Office 2018) 50 <https://fra.europa.eu/en/publication/2018/handbook-european-non-discrimination-law-2018-edition>. Jeremias Adams-Prassl, Reuben Binns and Aislinn Kelly-Lyth, 'Directly Discriminatory Algorithms' [2022] The Modern Law Review 1468 <https://onlinelibrary.wiley.com/doi/10.1111/1468-2230.12759>.
[25] Article 2(2)(b) of the Racial Equality Directive 2000/43/EC. The CJEU says that 'the concept of objective justification must be interpreted strictly'. *Nikolova* [2015] CJEU (Grand chamber) Case C-83/14, ECLI:EU:C:2015:480 [112].
[26] See in more detail about on applying EU non-discrimination law to AI-driven discrimination: F Zuiderveen Borgesius, 'Price discrimination, algorithmic decision-making, and European non-discrimination law'. European Business Law Review, 2020, 31.3.
[27] Article 8 of the European Convention on Human Rights.
[28] Article 17 of the International Covenant on Civil and Political Rights.
[29] Article 7 of the Charter of Fundamental Rights of the European Union.
[30] See M Granger and K Irion, 'The Right to Protection of Personal Data: The New Posterchild of European Union Citizenship?' in S de Vries, H de Waele and M Granger, *Civil Rights and EU Cit-*

## 5. Does the GDPR hinder the prevention of discrimination?

### 5.1. *The GDPR's ban of processing special categories of data*

With specific rules, the EU General Data Protection Regulation (GDPR) further works out the right to data protection from the EU Charter. The GDPR, like its predecessor, the Data Protection Directive from 1995, contains an in-principle prohibition on using certain types of extra sensitive data, called 'special categories' of data.[32] Some older data protection instruments also included stricter rules for special categories of data. Data protection law's stricter regime for special categories of data can be explained, in part, by the wish to prevent unfair discrimination. In 1972, the Council of Europe said about sensitive personal data: 'In general, information relating to the intimate private life of persons or information which might lead to unfair discrimination should not be recorded or, if recorded, should not be disseminated'.[33] The Guidelines for the Regulation of Computerized Personal Data Files of the United Nations (1990) use the title 'principle of non-discrimination' for its provision on special categories of data.[34]

The GDPR also refers to the risk of discrimination in its preamble. Recital 71 concerns AI and calls upon organisations[35] to 'prevent, inter alia, discriminatory effects on natural persons on the basis of racial or ethnic origin, political opinion, (…) or that result in measures having such an effect.'[36]

Article 9(1) GDPR is phrased as follows:

> Processing of personal data revealing racial or ethnic origin, political opinions, religious or philosophical beliefs, or trade union membership, and the processing of genetic data, biometric data for the purpose of uniquely identifying a natural person, data concerning health or data concerning a natural person's sex life or sexual orientation shall be prohibited.

Most protected grounds in EU non-discrimination directives are also special categories of data as defined in Article 9(1) GDPR. There are two exceptions. First, 'age' and 'gender' are protected characteristics in non-discrimination law, but are not special categories of data in the sense of the GDPR.[37] Second, 'political opinions', 'trade union membership', 'genetic' and 'biometric' data are special categories of data but are not protected by the European non-discrimination Directives.[38]

We summarize the distinction between the 'special categories of data' and 'protected non-discrimination grounds' in Fig. 1.

The GDPR's prohibition to process special category can hinder the prevention of discrimination by AI systems.[39] Think about the following scenario. An organisation uses an AI-driven recruitment system to select the best applicants from many job applications. The organisation wants to check whether its AI system accidentally discriminates against certain ethnic groups. For such an audit, the organisation requires data concerning the ethnicity of the job applicants. Without such data, it's very difficult to do such an audit.

However, Article 9(1) of the GPPR prohibits using such ethnicity data. Article 9(1) not only includes explicit information about a data subject's ethnicity, but also information 'revealing' ethnicity. Hence, the organisation is not allowed to infer the ethnicity of its applicants either.[40] The GDPR contains exceptions to the ban; we discuss those in the next section.

### 5.2. *The exceptions to the ban*

Article 9(2) GDPR includes a list of exceptions to the general prohibition to process special category data. The subsections (a), (b), (f), (g) and (j) contain possibly relevant exceptions for collecting special categories of data. The exceptions concern (a) explicit consent, and specific exceptions for (b) so-

---

*izenship* (Edward Elgar 2018) <https://www.elgaronline.com/view/edcoll/9781788113434/9781788113434.00019.xml>.

[31] Article 21 of the Charter of Fundamental Rights of the European Union.

[32] Article 8 of the Data Protection Directive.

[33] Committee of Ministers, Resolution (73)22 on the protection of the privacy of individuals vis-à-vis electronic data banks in the private sector, 26 September 1973, article 1. https://rm.coe.int/1680502830.

[34] Principle 5. https://www.refworld.org/docid/3ddcafaac.html.

[35] The GDPR puts most responsibilities on the 'data controller', in short, the body which determines the purposes and means of the personal data processing (Article 4(7) GDPR). For ease of reading, we speak of the 'organisation' in this paper.

[36] A hypothetical example might be as follows. An organisation uses AI to select the best candidates from a number of job application letters. Recital 71 reminds the organisation that it should prevent that its system discriminates unfairly, for instance on the basis of ethnicity.

[37] We add a caveat. In some circumstances, age and gender could be special categories of data if they were broadly interpreted under special categories 'health data', 'biometric data' or 'genetic data'. See M Van den Brink and P Dunne, *Trans and Intersex Equality Rights in Europe: A Comparative Analysis.* (Publications Office 2018) 9 <https://data.europa.eu/doi/10.2838/75428>.

[38] See European Union Agency for Fundamental Rights, European Court of Human Rights, and Council of Europe (Strasbourg) (n 24) s 5.11. We add a caveat: under circumstances, these special categories of data may strongly correlate with protected characteristics, directly revealing them.

[39] K Alidadi, 'Gauging Progress towards Equality? Challenges and Best Practices of Equality Data Collection in the EU' (2017) 2 European Equality Law review 21–22. Y Al-Zubaidi, 'Some Reflections on Racial and Ethnic Statistics for Anti-Discrimination Purposes in Europe' 2 European Equality Law review 2020, 65.

[40] Christopher Kuner and others (eds), *The EU General Data Protection Regulation (GDPR): A Commentary* (Oxford University Press 2020) s C.1 <https://oxford.universitypressscholarship.com/10.1093/oso/9780198826491.001.0001/isbn-9780198826491>.

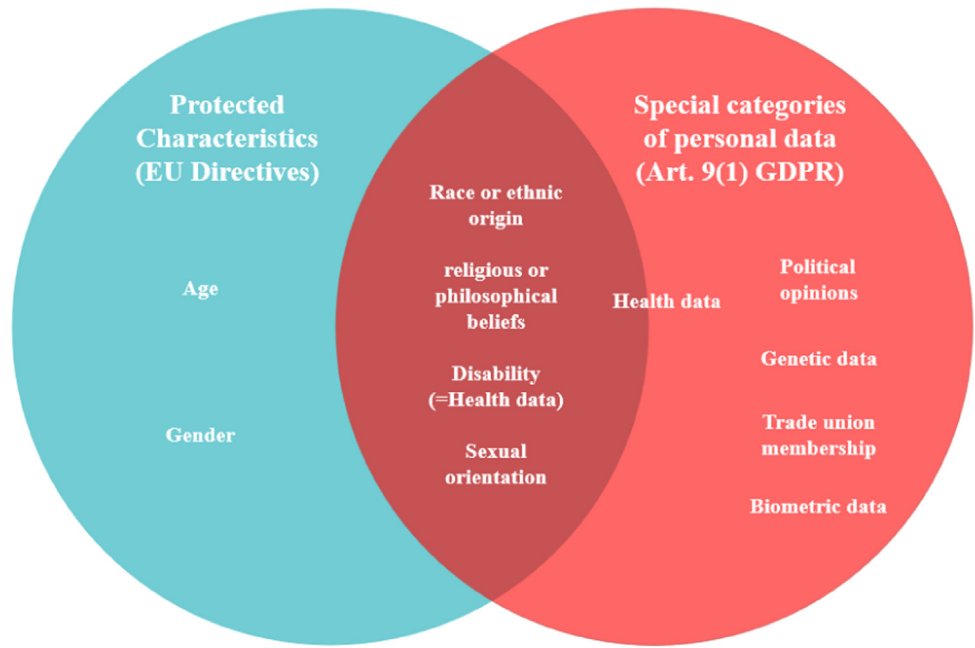

**Fig. 1 – The overlap between protected characteristics and special categories of data.**

cial security and social protection law, (f) legal claims, (g) reasons of substantial public interest, and (j) research purposes. Georgieva & Kuner say that all these exceptions 'are to be interpreted restrictively'.[41]

Some non-discrimination scholars appear to suggest that data protection law does not hinder collecting or using special categories of data to fight discrimination.[42] One scholar suggests that there is a 'need to "myth bust" the notion that data protection legislation should preclude the processing of equality data.'[43] However, we did not find detailed argumentation in the literature for the view that the GDPR allows using special categories of data for non-discrimination purposes. Below we show that, in most circumstances, the GDPR does not allow the use of special category data to fight discrimination.

### 5.3. Explicit consent

We discuss each exception to the ban on processing special category data in turn, starting with the data subject's consent. The ban does not apply if '(a) the data subject has given explicit consent to the processing of those personal data for one or more specified purposes (…).'[44]

In short, the data subject's explicit consent can lift the ban. However, the requirements for valid consent are very strict.[45] Valid consent requires that consent is 'specific' and 'informed', and requires an 'unambiguous indication of the data subject's wishes by which he or she, by a statement or by a clear affirmative action, signifies agreement to the processing of personal data relating to him or her.'[46] Hence, opt-out systems cannot be used to obtain valid consent.

Moreover, Article 4(11) GDPR prescribes that consent must be 'freely given' to be valid. Valid consent thus requires that the consent is voluntary. The GDPR's preamble gives some guidance on interpreting the 'freely given' requirement: 'Consent should not be regarded as freely given if the data subject has no genuine or free choice or is unable to refuse or withdraw consent without detriment.'[47]

Consent is less likely to be voluntary if there is an imbalance between the data subject and the controller. The preamble says that 'consent should not provide a valid legal ground for the processing of personal data in a specific case where there is a clear imbalance between the data subject and the

---

[41] ibid C.3. 'The list of exceptions is exhaustive and all of them are to be interpreted restrictively'.

[42] L Farkas, The Meaning of Racial or Ethnic Origin in EU Law: Between Stereotypes and Identities, European Network of Legal Experts in Gender Equality and Non-Discrimination (Report for European Commission, Directorate-General for Justice and Consumers) (Publications Office of the European Union 2017) 14 <https://www.equalitylaw.eu/downloads/4030-the-meaning-of-racial-or-ethinic-origin-in-eu-law-between-stereotypes-and-identities>. K. Alidadi, 'Gauging Progress towards Equality? Challenges and Best Practices of Equality Data Collection in the EU' (2017) 2 European Equality Law review 21–22.

[43] Alidadi (n 39). Al-Zubaidi (n 39) 22. See also T Makkonen, *European Handbook on Equality Data: 2016 Revision* (Publications Office 2016) 27 <https://data.europa.eu/doi/10.2838/397074>. L Farkas, *The Meaning of Racial or Ethnic Origin in EU Law: Between Stereotypes and Identities, European Network of Legal Experts in Gender Equality and Non-Discrimination (Report for European Commission, Directorate-General for Justice and Consumers)* (Publications Office of the European Union 2017) 14 <https://www.equalitylaw.eu/downloads/4030-the-meaning-of-racial-or-ethinic-origin-in-eu-law-between-stereotypes-and-identities>.

[44] Article 9(2)(a) GDP.
[45] See article 4(11) and 7 GDPR.
[46] Article 4(11) and 7 GDPR.
[47] Recital 43 GDPR.

controller'. If there is a 'clear' imbalance, consent is not freely given and thus invalid.'[48]

The European Data Protection Board says that consent from an employee to an employer is usually not valid:[49]

> An imbalance of power also occurs in the employment context. Given the dependency that results from the employer/employee relationship, it is unlikely that the data subject is able to deny his/her employer consent to data processing without experiencing the fear or real risk of detrimental effects as a result of a refusal.[50]

The Board adds:

> [T]he EDPB deems it problematic for employers to process personal data of current or future employees on the basis of consent as it is unlikely to be freely given. For the majority of such data processing at work, the lawful basis cannot and should not be the consent of the employees (Article 6(1)(a)) due to the nature of the relationship between employer and employee.[51]

What does this mean for using special category data for preventing AI-driven discrimination? In many situations this 'freely given' requirement poses a problem. We return to our example: an organisation uses AI to select the best job applicants, and wants to audit its AI system for accidental discrimination. The organisation might consider asking all job applicants for consent to collect data about their ethnicity, to use that information for auditing its AI system. However, as discussed, applicants could fear that they would be rejected because of refusing to share their special category data. The applicant's consent is therefore generally invalid.

Perhaps a system could be designed in which a job applicant can give genuinely 'freely given' and thus valid consent. For instance, an organisation could ask all rejected job applicants for their consent after the position has been filled. In that case, job applicants might not fear anymore that withholding consent diminishes their chances to get the job. However, the organisation might find it awkward to ask people about their ethnicity, religion, or sexual preferences. Moreover, people can refuse their consent. If too many people refuse, the sample will not be representative – and cannot be used to audit AI systems.

We gave one specific example where gathering consent would be non-compliant. In many other contexts, particularly those where no power relationship exists between data subject and data controller, the data subject's consent may provide an exception to collect data regarding e.g. ethnicity.[52] All in all, in certain circumstances, organisations could rely on consent to lift the ban. But in many circumstances, the data subject's consent seems not a real possibility.

### 5.4.　*Other exceptions*

We continue with the other exceptions to the ban on processing special category data, starting with (b), on employment and social security and social protection law. The ban can be lifted if:

> (b) processing is necessary for the purposes of carrying out the obligations and exercising specific rights of the controller or of the data subject in the field of employment and social security and social protection law (…) in so far as it is authorised by Union or Member State law or a collective agreement pursuant to Member State law.[53]

Exception (b) only applies to situations relating to the field of employment and social security and social protection law. An example is a provision of the Dutch implementation legislation of the GDPR.[54]

Roughly summarized, that provision allows employers to collect health data of employees if that is necessary for e.g. their re-integration after an illness. In a case about this provision, the Dutch Data Protection Authority stated that an employer must regularly re-consider if gathering the health data is still truly *necessary* in the light of the employee's re-integration duty. This requirement followed from a restrictive interpretation of Art. 9(1)(b) GDPR.[55]

Could this GDPR provision help organisations that want to audit their AI systems for discrimination? No. The main problem is that exception (b) only applies if the EU lawmaker or the national lawmaker has adopted a specific law that enables the use of special category data. To the best of our knowledge, no national lawmaker in the EU, nor the EU, has adopted a specific law that enables the use of special category data for auditing AI systems.[56] We turn to the next possibly relevant exception in the GDPR: the ban does not apply if

---

[48] See on an imbalance between a data controller and a data subject also: Eleni Kosta, 'Unravelling Consent in European Data Protection Legislation - a Prospective Study on Consent in Electronic Communications' (KU Leuven 2011) 178–183 <https://limo.libis.be/primo-explore/fulldisplay?docid=LIRIAS1710851>.

[49] The EDPB is an advisory body in which all European Data Protection Authorities cooperate. The body's opinions are considered persuasive in data protection law, carrying some weight.

[50] European Data Protection Board, *Guidelines 05/2020 on Consent under Regulation 2016/679* (EDPB 2020) 9 <https://edpb.europa.eu/sites/default/files/files/file1/edpb_guidelines_202005_consent_en.pdf>.

[51] ibid.

[52] Makonnen writes: 'it is subparagraph a on the consent of the data subject which is likely to become the most frequently used basis for processing sensitive data.' Makkonen (n 43) 28.

[53] Article 9(2)(b) GDPR.

[54] Art. 30 *Uitvoeringswet Algemene Verordening Gegevensbescherming* (UAVG): 'In view of Article 9, Section 2, under b, of [the GDPR], the prohibition on processing data concerning health does not apply if the processing is carried out by administrative bodies, pension funds, employers or institutions working on their behalf, and in so far as the processing is necessary for: (a) the proper execution of statutory provisions, pension schemes or collective agreements providing for entitlements which depend on the state of health of the person concerned; or (b) the reintegration or assistance of employees or beneficiaries in connection with illness or disability.' Translation by the authors of this paper.

[55] See in Dutch: Dutch Data Protection Authority, *Besluit Tot Het Opleggen van Een Bestuurlijke Boete [Decision to Impose an Administrative Fine]* (2020) 7 <https://autoriteitpersoonsgegevens.nl/nl/nieuws/boete-voor-cpa-om-privacyschending-zieke-werknemers>.

[56] A specific duty may be present in Finnish law. We do not know if this duty was intended for auditing AI systems. The duty exists for

(f) processing is necessary for the establishment, exercise or defence of legal claims or whenever courts are acting in their judicial capacity.

Exception (f) applies to 'legal claims'; hence, claims in legal proceedings such as court cases. The exception applies to the use of personal data by courts themselves. An organisation might argue that it needs to audit its AI systems to prevent future lawsuits because of illegal discrimination. However, Kuner & Bygrave note that exception (f) is only valid for concrete court cases. The exception 'does not apply when sensitive data are processed merely in anticipation of a potential dispute without a formal claim having been asserted or filed, or without any indication that a formal claim is imminent.'[57]

It seems implausible that an organisation can use it to collect special categories of data about many people to audit its AI systems preventatively. Could exception (g) and (j) help organisations that want to debias their AI systems?

(g) processing is necessary for reasons of substantial public interest, on the basis of Union or Member State law (…).

(j) processing is necessary for archiving purposes in the public interest, scientific or historical research purposes or statistical purposes in accordance with Article 89(1) based on Union or Member State law (…).

The two provisions require a legal basis in a law of the EU or of a national lawmaker.[58] Hence, the provisions do not allow processing of special categories of data. Instead, exceptions (g) and (j) give flexibility to the EU or member states to decide whether processing special category data for fighting discrimination is allowed.[59] Current EU law, nor national law, provide such an exception.

To fight discrimination under exceptions (g) and (j), national law must provide a legal ground for processing. To the best of our knowledge, the United Kingdom was the only member state to have ever adopted an exception. (Now the UK is not an EU member anymore). The UK exception is based on Article 9(2)(g), the 'substantial public interest' exception. [60] However, the EU nor the current member states have adopted such a law. (The proposed AI Act contains such a provision; we discuss that in Section 7.1).

During the drafting of the GDPR, the Fundamental Rights Agency (FRA), a European Union agency, realised that the (proposed) GDPR did not allow such data collection. The Agency said that '[the GDPR] could clarify the place of special category categories in anti-discrimination data collection, and make explicit that the collection of special category data is allowed for the purpose of combatting discrimination'.[61] But the EU did not follow that suggestion.

For completeness' sake, we highlight that an organisation is not out of the woods yet if it has found a way to lift the ban on using special category data. For the processing of personal data, sensitive or not, the GDPR requires a 'legal ground'. Hence, even if the ban on using special category data could be lifted, the organisation must still find a valid legal processing ground as defined in Article 6(1) GDPR. For non-state actors, the legitimate interest ground (Article 6(1)(f) GDPR) seems the most plausible. The organisation must also comply with all the other requirements of the GDPR. But a discussion of all those GDPR requirements falls outside the scope of this paper.

In conclusion, the GDPR indeed hinders organisations who wish to use special category data to prevent discrimination by their AI systems. In some exceptional situations, an organisation might be able to obtain valid consent from data subjects for such use. In other situations, an EU or national law would be needed to enable the use of special categories of data for AI debiasing – at the moment such laws are not in force in the EU. In the next section, we explore whether such an exception is a good idea.

## 6. A new exception to the ban on using special categories of data?

### 6.1. Introduction

Policymakers have realised that non-discrimination policy can conflict with data protection law, and some have adopted exceptions. As noted, the UK has adopted an exception to the ban on using special categories of data, for the purpose of fighting discrimination.[62] The Dutch government is considering to create a new national exception for collecting special categories of data.[63] We are aware of 6 countries out-

---

'all authorities, all employers and all providers of education' to 'assess the realisation of equality in their functions'. Farkas further notes: 'Obligations to collect racial and ethnic data do not generally seem to be codified in law in the Member States.' See Farkas (n 43) 15.

[57] Kuner and others (n 40) s 3, summation, under 6.

[58] Personal data collection for equality and anti-discrimination purposes likely constitutes a substantial public interest. See J Van Caeneghem, 'Ethnic Data Collection: Key Elements, Rules and Principles' in J Van Caeneghem, *Legal Aspects of Ethnic Data Collection and Positive Action* (Springer International Publishing 2019) 215, 217, 219, 242 <http://link.springer.com/10.1007/978-3-030-23668-7_3>.

[59] ibid 218.

[60] The UK data protection Act 2018 itself states 'substantial public interest' as the legal ground. See UK Data Protection Act 2018, Schedule 1 Part 2 https://www.legislation.gov.uk/ukpga/2018/12/schedule/1/part/2/crossheading/equality-of-opportunity-or-treatment. See also ICO, 'What Are the Substantial Public Interest Conditions?' (8 June 2021) <https://ico.org.uk/for-organisations/guide-to-data-protection/guide-to-the-general-data-protection-regulation-gdpr/special-category-data/what-are-the-substantial-public-interest-conditions/> accessed 12 April 2022.

[61] FRA, *FRA Opinion on the Situation of Equality in the European Union 10 Years on from Initial Implementation of the Equality Directives* (2013) 21–22 <https://fra.europa.eu/en/publication/2013/fra-opinion-situation-equality-european-union-10-years-initial-implementation>.

[62] UK Data Protection Act 2018, Schedule 1 Part 2 https://www.legislation.gov.uk/ukpga/2018/12/schedule/1/part/2/crossheading/equality-of-opportunity-or-treatment. See ICO website, ICO (n 60). See also Open Data Institute, 'Monitoring Equality in Digital Public Services' (2020) 17 <https://theodi.org/wp-content/uploads/2020/01/OPEN-ODI-2020-01_Monitoring-Equality-in-Digital-Public-Services-report.pdf>.

[63] See in Dutch answer to question 12 of https://zoek.officielebekendmakingen.nl/kst-26643-727.html and par. 2.2 of

side the EU with an exception in their national data protection act that enables the use of special category data for non-discrimination purposes. These countries are Bahrain, Curaçao, Ghana, Jersey, Sint Maarten, and South Africa.[64] The European Commission presented the AI Act proposal, with a possible new exception in Article 10(5) AI Act (see Section 7.1).

In the following section, we map out arguments in favour and against creating an exception for gathering special categories of data for the purpose of auditing an AI system for discrimination.

### 6.2. Arguments in favour of an exception

We present two main arguments in favour of creating a new exception that enables the use of special category data to prevent AI-driven discrimination: (i) Several types of organisations could use the data to test whether an AI system discriminates. (ii) The collection of the data has a symbolic function.

A first, rather strong, argument is that, for many types of stakeholders, collecting special category data would make the fight against discrimination easier. Organisations could check, themselves, whether their AI system accidentally discriminates. Organisations may want to ensure that their hiring, firing and other policies and practices comply with non-discrimination laws. Organisations may care about fairness and non-discrimination, or may want to protect their reputation.[65]

Regulators, such as equality bodies (non-discrimination authorities) could also benefit from an exception that enables the use of special categories of data for AI debiasing. Regulators could more easily check an organisation's AI practices if those organisations registered the ethnicity of all their employees, job applicants, etc.[66]

Another group that can benefit from the collection of special category data is researchers.[67] Researchers could use such data to check whether an AI system discriminates. This argument is only valid, however, if an organisation shares its data with the researcher.

A different type of argument in favour of allowing the use of special category data is related to the symbolic function of such use.[68] Auditing an AI system can increase the trust in an organisation's AI practices, if it is publicly known that the organisation checks whether its AI systems discriminate. Potential discrimination victims can see that the organisation takes discrimination by AI seriously.[69] We mention this argument for completeness' sake. However, we do not see this as a particularly strong argument. In sum, there are various arguments in favour of adopting an exception that enables the use of special category data to prevent discrimination by AI.

### 6.3. Arguments against an exception

There are also strong arguments against adopting an exception that enables the use of special category data for AI debiasing. Balayn & Gürses warn for the danger of surveillance of protected groups: 'Policy that promotes debiasing (…) may incentivise increased data collection of exactly those populations who may be vulnerable to surveillance.'[70]

We distinguish three categories of arguments against introducing a new exception to enable the use of special categories of data to mitigate discrimination risks. There are arguments (i) that concern the mere storage of special categories of data, (ii) that concern new uses of those data, and (iii) that show practical hurdles as a reason why an exception is not justified at this time.

We start with arguments that relate to the sole fact that an organisation stores special category data, regardless of how the data are used. People may feel uneasy if their data are collected or stored. People may have that feeling, regardless of whether the data are accessed by anyone. Many people are uncomfortable with organisations storing large amounts of personal data about them, even if no human ever looks at the data. Calo speaks of subjective harm, 'the perception of loss of

---

https://zoek.officielebekendmakingen.nl/kst-26643-726.html.' See p. 3, final paragraph of https://www.rijksoverheid.nl/documenten/kamerstukken/2020/12/04/tk-reactie-op-mededelingen-europese-commissie-over-de-avg.
[64] Bahrain Personal Data Protection Act of 2018, Article 5 www.legalaffairs.gov.bh/146182.aspx?cms=q8FmFJgiscJUAh5wTFxPQnjc67hw%2bcd53dCDU8XkwhyDqZn9xoYKj%2bwKjH8MwskD8zKV4oL8QNchAeJU7Z6zGg%3d%3d#.XAAo1NtKiUl. https://eur-lex.europa.eu/legal-content/EN/TXT/?uri=CELEX%3A52021PC0206. Curacao National Ordinance Protection of Personal Data - Chapter 2(2), https://media2.mofo.com/documents/Curacao-PRIVACY-ACT.pdf. Ghana Data Protection Act, 2012, https://www.dataprotection.org.gh/data-protection/data-protection-acts-2012. Jersey https://www.dataprotection.org.gh/data-protection/data-protection-acts-2012, Schedule 2 Pt. 1, https://www.jerseylaw.je/laws/enacted/PDFs/L-03-2018.pdf. South Africa Protection of Personal Information Act 2013 Ch. 3, www.saflii.org/za/legis/num_act/popia2013380.pdf. Saint Martin Personal Data Protection Act 2010 GT No. 2 Chapter 2 Par. 2, http://www.sintmaartengov.org/government/AZ/laws/AFKONDIGINGSBLAD/AB%2002%20Landsverordening%20bescherming%20persoonsgegevens.pdf.
[65] See also Makkonen (n 43) 21.
[66] Makkonen: 'National specialised bodies, such as ombudsmen and equality bodies, and international monitoring bodies, such as the UN treaty bodies and the Council of Europe's European Commission against Racism and Intolerance (ECRI), as well as some other institutions, such as the EU Fundamental Rights Agency,

need quantitative and qualitative information in order to perform their functions properly.' ibid 20.
[67] ibid 21.
[68] Similar to an argument made by Makonnen: '[…] the compilation of equality statistics can be seen to have more symbolic functions. The mere existence of a data collection system sends a message to actual and potential perpetrators, actual and potential victims and to society in general, signalling that society disapproves of discrimination, takes it seriously and is willing to take the steps necessary to fight it. This can have a preventive effect.' ibid.
[69] See for a similar argument for collecting non-discrimination data in general Alidadi (n 39) 18.
[70] Agathe Balayn and Seda Gürses, *Beyond Debiasing. Regulating AI and Its Inequalities* (European Digital Rights (EDRi) 2021) 94 <https://edri.org/our-work/if-ai-is-the-problem-is-debiasing-the-solution/>.

control that results in fear or discomfort.'[71] Such harms could also be called 'expected harms'.[72]

The Court of Justice of the European Union accepts that the mere storage of personal data can interfere with the rights to privacy and to the protection of personal data.[73] The European Court of Human Rights has also said that storing sensitive personal data can interfere with the right to privacy, regardless of how those data are used: 'even the mere storing of data relating to the private life of an individual amounts to an interference within the meaning of Article 8' of the European Convention on Human Rights, which protects the right to privacy.[74]

In sum, the two most important courts in Europe accept that merely storing personal data can interfere with fundamental rights. Indeed, we think that many people may dislike it when special category data about them (such as their ethnicity) are stored.

(i) A second category of arguments concerns risks related to the storage of data. Data that have been collected can be used for many new purposes, that can harm both the individuals involved and society as a whole. These arguments relate to 'experienced harm' or 'objective harm'.[75]

For instance, a data breach can occur. Storing data always brings the risk of data breaches. Employees at an organisation or outsiders may gain unauthorised access to the data. A breached database with personal data about people's ethnicity, religion, or sexual preferences could have very negative effects. From a data security perspective, it is good policy not to develop databases with such sensitive data.

Second, there is a risk that data are used for new, unforeseen, uses. An extreme example of a new use of data about ethnicity and religion concerns the registration of Jewish people in the Netherlands, during the time that the Nazis occupied the Netherlands. The Nazis could easily find Jewish people in the Netherlands, because the citizen registration included the ethnicity and religion for each person. The IBM computers in the citizen registry made it easy to compile lists of, for instance all the Jewish people in Amsterdam.[76] There are many more examples where collecting population data facilitated human rights abuses.[77] Seltzer and Anderson write: 'In many large-scale human rights abuses, statistical outputs, systems, and methods are a necessary part of the effort to define, find, and attack an initially dispersed target population.'[78]

Third, organisations could misuse the exception to collect large amounts of special categories of data, claiming that they need such data to fight discrimination. An exception that is too wide could open the door for mass data gathering. As Balayn and Gürses note, 'the possibility that debiasing methods may lead to over-surveillance of marginalised populations should be a very serious concern.'[79]

Fourth, a symbolic argument can be made against an exception that allows the collection and storage of special category data. People want their sensitive data to be handled with care. If people know that an organisation does not collect their special category data, they could trust that organisation more. (As with the symbolic argument in favour of using special category data, we do not think this argument is very strong.) In sum, there are several arguments against adopting an exception that enables using special categories of personal data to prevent discrimination by AI.

Finally, there is a different category of arguments against adopting a new exception to enable collecting and using special categories of data for AI non-discrimination auditing. The world seems not yet ready for such an exception, as auditing AI systems is still very difficult.

Andrus, Spitzer, Brown and Xiang note that 'the algorithmic fairness literature provides a number of techniques for potentially mitigating detected bias, but practitioners noted that there are few applicable, touchstone examples of these techniques in practice.'[80] Balayn and Gürses write: '[w]hether applied to data-sets or algorithms, technocentric debiasing techniques have profound limitations: they address bias in mere statistical terms, instead of accounting for the varied and complex fairness requirements and needs of diverse system stakeholders.'[81]

A 2021 interview study found several practical reasons explaining why industry practitioners *themselves* find it difficult to test if an AI system discriminates. For instance, the collected special category data may not be accurate, because the data may have originally been collected by another party, or for a different purpose. And because data subjects may report their own (self-perceived) ethnicities, the data could be inaccurate or unusable for the test.[82] An interview study from 2019 found that many industry practitioners did not have an infrastructure in place for collecting accurate special categories of data. Furthermore, practitioners mentioned that auditing methods themselves are not holistic enough, and practitioners do not always know which subpopulations they need to consider when auditing an AI system for discrimination.[83]

---

[71] MR Calo, 'The Boundaries of Privacy Harm' 86 Indiana Law Journal 31, 1143.

[72] Fahriye Seda Gurses, 'Multilateral Privacy Requirements Analysis in Online Social Network Services' 312, 87–89.

[73] CJEU, C-291/12, Schwartz v. Stadt Bochum, 17 October 2013, par. 25: 'as a general rule, any processing of personal data by a third party may constitute a threat to those rights'. See also the judgment on the Data Retention Directive: CJEU, C-293/12 and C-594/12, Digital Rights Ireland Ltd, 8 April 2014, par. 29.

[74] ECtHR (Grand Chamber), Big Brother Watch and others v. the United Kingdom, No. 58170/13, 62322/14 and 24960/15, 25 May 2021, Par. 330. See also: Leander v. Sweden, 26 March 1987, § 48, Series A no. 116; ECtHR, Copland v. United Kingdom, No. 62617/00, 3 April 2007, par. 43-44; ECtHR, Amann v. Switzerland, No. 27798/95, 16 February 2000, par. 69; ECtHR, S. and Marper v. United Kingdom, No. 30562/04 and 30566/04. 4 December 2008, par. 67, par 121.

[75] Calo describes objective harm as 'the unanticipated or coerced use of information concerning a person against that person.' Calo (n 71) 1133. Gurses (n 72) 87–89.

[76] E. Black, *IBM and the Holocaust: The Strategic Alliance Between Nazi Germany and America's Most Powerful Corporation* (Dialog press 2012).

[77] For 10 cases, see See Table 1 of W Seltzer and M Anderson, 'The Dark Side of Numbers: The Role of Population Data Systems in Human Rights Abuses' <https://www.jstor.org/stable/40971467>.

[78] ibid 507.

[79] Balayn and Gürses (n 70) 94.

[80] M Andrus and others, '"What We Can't Measure, We Can't Understand": Challenges to Demographic Data Procurement in the Pursuit of Fairness' [2021] arXiv:2011.02282 [cs] 257 <http://arxiv.org/abs/2011.02282>.

[81] Balayn and Gürses (n 70) 12. See also M Veale and R Binns, 'Fairer Machine Learning in the Real World: Mitigating Discrimination without Collecting Sensitive Data' (2017) 4 Big Data & Society 205395171774353, 13 <http://journals.sagepub.com/doi/10.1177/2053951717743530>: '[t]o adequately address fairness in the context of machine learning, researchers and practitioners working towards "fairer" machine learning need to recognise that this is not just an abstract constrained optimisation problem. It is a messy, contextually-embedded and necessarily sociotechnical problem, and needs to be treated as such.'

Since 'best practices' for auditing an AI system for discrimination seem to be in their infancy, it is questionable whether creating a new exception is currently justified. However, the techniques for auditing and debiasing AI are improving. The more knowledge exists about how to test an AI system for discrimination, the more justified a new exception could become in the future.

All in all, there are various arguments in favour of and against adopting an exception that enables the use of special categories of data to prevent AI-driven discrimination. The balance between the pro and contra arguments is difficult to find. If such an exception were adopted, the exception should also include safeguards to minimise risks. We discuss possible safeguards in the next section.

## 7. Possible safeguards if an exception were adopted

### 7.1. Safeguards in the proposed AI Act

A proposal by the EU illustrates some possibilities for safeguards. Early 2021, the European Commission presented a proposal for an AI Act, with an exception to the ban on using special category data. The proposed exception is phrased as follows:[84]

> To the extent that it is strictly necessary for the purposes of ensuring bias monitoring, detection and correction in relation to the high-risk AI systems, the providers of such systems may process special categories of personal data referred to in [Article 9 of the GDPR], subject to appropriate safeguards for the fundamental rights and freedoms of natural persons, including technical limitations on the re-use and use of state-of-the-art security and privacy-preserving measures, such as pseudonymisation, or encryption where anonymisation may significantly affect the purpose pursued.

There are various safeguards in the proposed provision that aim to prevent abuse of the special category data.

#### 7.1.1. Strictly necessary for preventing discrimination
In the AI Act, the exception to the ban on using special category data only applies '[t]o the extent that it is strictly necessary for the purposes of ensuring bias monitoring, detection and correction in relation to the high-risk AI systems'.

The phrase 'strictly necessary' implies a higher bar than merely 'necessary'. The word 'necessary' is already quite stern. The CJEU said in the *Huber* case that 'the concept of necessity (…) has its own independent meaning in Community law.'[85] And necessity 'must be interpreted in a manner which fully reflects the objective of [the Data Protection] directive'.[86] CJEU case law shows the word 'necessary' must be interpreted narrowly, in favour of the data subject: 'As regards the condition relating to the necessity of processing personal data, it should be borne in mind that *derogations and limitations in relation to the protection of personal data must apply only in so far as is strictly necessary* (…).'[87] In sum, organisations should only rely on the exception in the AI Act if using special category data is genuinely necessary.

#### 7.1.2. The AI Act exception only applies to providers of high-risk AI systems
The AI Act's exception only applies to high-risk AI systems. High-risk AI systems can be divided into two types: First, products already covered by certain EU health and safety harmonisation legislation (such as toys, machinery, lifts, or medical devices). Second, AI systems specified in an annex of the AI Act, in eight areas or sectors.[88]

If lawmakers consider creating a new exception to enable the use of special category data, they could limit the scope of the exception in a similar way. Perhaps organisations should not be allowed to rely on the exception if the AI system does not bring serious discrimination risks.

#### 7.1.3. Appropriate safeguards
The exception in the proposed AI Act says that special category data can be used to prevent AI-driven discrimination,

---

[82] Andrus and others (n 80) s 6.3.
[83] Kenneth Holstein and others, 'Improving Fairness in Machine Learning Systems: What Do Industry Practitioners Need?' [2019] Proceedings of the 2019 CHI Conference on Human Factors in Computing Systems 1 <http://arxiv.org/abs/1812.05239>.
[84] Quotation lightly edited by authors. See Article 10(5) and recital 44 of the European Commission Proposal for a Regulation of the European Parliament and of the Council laying down harmonised rules on Artificial Intelligence (Artificial Intelligence Act) and amending certain Union legislative acts. COM/2021/206 final. https://eur-lex.europa.eu/legal-content/EN/TXT/?uri=CELEX%3A52021PC0206.
[85] ECJ, Case C-524/06 Huber [2008] ECLI:EU:C:2008:724, par. 52.
[86] ECJ, Case C-524/06 Huber [2008] ECLI:EU:C:2008:724, par. 52
[87] CJEU, C–13/16, Rigas, 4 May 2017, par. 30: 'As regards the condition relating to the necessity of processing personal data, it should be borne in mind that derogations and limitations in relation to the protection of personal data must apply only in so far as is strictly necessary.' See also CJEU, C-293/12 and C-594/12, Digital Rights Ireland Ltd, 8 April 2014, par. 52: 'according to the Court's settled case-law, (…) derogations and limitations in relation to the protection of personal data must apply only in so far as is *strictly necessary*.' See also Silver and Others v United Kingdom App no 5947/72; 6205/73; 7052/75; 7061/75; 7107/75; 7113/75; 7136/75 (ECHR, 25 March 1983), par 97: The European Court of Human Rights says that '[t]he adjective 'necessary' is not synonymous with 'indispensable', neither has it the flexibility of such expressions as 'admissible', 'ordinary', 'useful', 'reasonable' or 'desirable' (…).' As stated previously in Section 5.4, the Dutch DPA has also used the phrase 'truly necessary', referring to a higher standard of proportionality.
[88] See M Veale and FJ Zuiderveen Borgesius, 'Demystifying the Draft EU Artificial Intelligence Act' (2021) 4 Computer Law Review International 102.

'subject to appropriate safeguards for the fundamental rights and freedoms of natural persons'.[89] The provision gives examples of such safeguards: 'including technical limitations on the re-use and use of state-of-the-art security and privacy-preserving measures, such as pseudonymisation, or encryption where anonymisation may significantly affect the purpose pursued.'[90] If an exception were adopted to enable the use of special category data for fighting AI-driven discrimination, a similar requirement should be included.

Some elements in the proposed AI Act exception are controversial. For instance, the current exception leaves unclear *who* decides what the appropriate safeguards are. With the current text, that burden seems to rest on the provider of the AI system him or herself. Therefore, the exception seems to leave the exact safeguards up to the provider of the AI system.

Moreover, as Balayn & Gürses note, the 'European Commission proposal to regulate AI enables the use of sensitive attributes for debiasing, without further consideration of the risks it imposes on exactly the populations that the regulation says it intends to protect.'[91] Indeed, the AI Act does not include measures to limit the risks associated with collecting special category data.

### 7.2. Other possible safeguards

Are other safeguards, not mentioned in the AI Act, viable? A specific technical safeguard is to create a synthetic, anonymous dataset from the 'real' dataset. The fake dataset represents the same (or a similar) distribution of individuals but can no longer be linked to the individuals. Such data can be safely stored.

The original special categories of data still need to be collected to create a synthetic dataset, but the original data can be stored for less time. It is controversial whether and when using such a dataset is effective for testing AI systems. At the current time of writing, the privacy gain seems to vary greatly, and it is unpredictable how much utility is lost by making the dataset synthetic.[92]

Other possible safeguards are more organisational. For example, a trusted third party could collect the special categories of data, store them and use them for auditing the AI system. The organisation using the AI system then no longer needs to store the data itself. However, it seems debatable if such a construction is practical, and financially feasible.[93]

Such organisational safeguards raise many questions. For example, which third party can be trusted with special categories of data? One possibility might be the national Statistics Bureaus of member states. Such bureaus could store and manage the special category data safely. In Europe, statistics bureaus have long been responsible for collecting special categories of data for statistical purposes, on a large scale. Or perhaps an independent supervising authority could appoint trustworthy researchers and give them access to the special categories of data, to prevent discrimination by AI systems. Specific privacy-related solutions also exist that might allow a third party to more safely use the data.[94]

### 8. Conclusion

In this paper, we examined whether the GDPR needs a new exception to the ban on using special categories of data, such that an organisation can mitigate discrimination by artificial intelligence. We mapped out the arguments in favour of and against such a new exception.

We presented the following main arguments in favour of such an exception. (i) Organisations could use the special category data to test AI against discrimination. (ii) AI discrimination testing could increase the trust consumers have in an AI system.

The main arguments against such an exception are as follows. (i) Storing personal data about, for instance, ethnicity can be seen as a privacy interference. (ii) Such data can be abused, or data breaches can occur. (iii) The exception could be abused to collect special categories of data for other uses than AI discrimination testing. (iv) In addition, merely allowing organisations to collect special category data does not guarantee that organisations can debias their AI systems. Auditing and debiasing AI systems remains difficult.

In the end, it is a political decision how the balance between the different interests must be struck. Ideally, such a decision is made after a thorough debate. We hope to inform that debate.

### Declaration of Competing Interest

The author(s) declared no potential conflicts of interest with respect to the research, authorship, and/or publication of this article.

### Data Availability

No data was used for the research described in the article.

---

[89] Article 10(3) Draft AI Act.
[90] Article 10(3) Draft AI Act.
[91] Balayn and Gürses (n 70) 94.
[92] Theresa Stadler, Bristena Oprisanu and Carmela Troncoso, 'Synthetic Data – Anonymisation Groundhog Day', *31st USENIX security symposium* (2022) <https://www.usenix.org/conference/usenixsecurity22/presentation/stadler>. See also on the limits of using synthetic datasets Steven M Bellovin, Preetam K Dutta and Nathan Reitinger, 'Privacy and Synthetic Datasets' 22 Stan. Tech. L. Rev. 1 (2019) 14 and further <https://law.stanford.edu/wp-content/uploads/2019/01/Bellovin_20190129.pdf>.
[93] See also Niki Kilbertus and others, 'Blind Justice: Fairness with Encrypted Sensitive Attributes' [2018] arXiv:1806.03281 [cs, stat] 1–2 <http://arxiv.org/abs/1806.03281>.

[94] To illustrate, say a statistics bureau audits an organisation's patented AI system. If both organisations use secure multi party computation (MPC), the statistics bureau cannot see the model of the organisation it audits and the organisation cannot see the special category data used for the audit, but they could audit the AI system together. A caveat is that MPC creates some overhead in calculations and is complex to set up. See ibid 3.2. See also Wirawan Agahari, Hosea Ofe and Mark de Reuver, 'It Is Not (Only) about Privacy: How Multi-Party Computation Redefines Control, Trust, and Risk in Data Sharing' [2022] Electronic Markets <https://link.springer.com/10.1007/s12525-022-00572-w>.